\documentstyle[graphicx,multicol,prl,aps,epsf]{revtex}
\begin{document}
\author{Konstantin Kikoin and Yshai Avishai}

\address{Ilse Katz Center for Nanotechnology and Department of Physics \\
Ben-Gurion University of the Negev, Beer Sheva 84105 Beer-Sheva, Israel}

\title{Double quantum dot as a spin rotator}
\maketitle

\begin{abstract}
It is shown that the low-energy spin states of double quantum dots (DQD)
with even electron occupation number $N$ possess the symmetry $SO(4)$
similar to that of a rigid rotator familiar in quantum mechanics (rotational
spectra of H$_2$ molecule, electron in Coulomb field, etc). The "hidden
symmetry" of the rotator manifests itself in the tunneling properties of the
DQD. In particular, the Kondo resonance may arise under asymmetric gate
voltage in spite of the even electron occupation of the DQD. Various
symmetry properties of spin rotator in the context of the Kondo effect are
discussed and experimental realization of this unusual scenario is proposed.\\
 \
\end{abstract}


\begin{multicols} {2}
\section{Introduction}

During recent years, the physics of single electron tunneling through a
quantum dot (QD) under the conditions of strong Coulomb blockade has been at
the focus of intense investigation\cite{Kouw97}. The number of electrons ${N}
$ in a dot can be regulated by a suitable gate voltage $V_{g}$ applied to an
electrode coupled capacitively to the dot. The Coulomb blockade suppresses
the tunneling through the dot unless the resonance between its energies
filled by $N$ and $N+1$ electrons occurs at certain values of $V_{g}$, when
it compensates the charging energy, i.e. ${\cal E}(N+1,V_{g})\approx {\cal E}%
(N,V_{g})$. The differential conductance $dI/dV_{sd}$ of a QD forms a
diamond-like patterns in the plan $(V_{sd},V_{g})$ where the non-conducting
''windows'' are separated by a network of Coulomb resonance lines (here $%
V_{sd}$ is the source-drain voltage).

Accurate low-temperature experiments demonstrated the existence of Kondo
resonances in the windows corresponding to {\it odd} occupation of the dot 
\cite{Gogo99} (O-diamonds). These resonances are seen as zero-bias anomalies
(ZBA), i.e. as bridges of finite conductance connecting two opposite
vertices of O-diamond-shape window at $V_{sd}\to 0.$ Besides, it was
predicted theoretically \cite{APK} and observed experimentally \cite{Cobd00}
that Kondo resonances can appear also in the even occupation windows
(E-diamonds) at strong enough magnetic fields. This unconventional magnetic
field induced Kondo effect arises because the spectrum of the dot possesses
a low-lying triplet excitation when the electron at the highest occupied
level is excited with spin flip. The Zeeman energy compensates the energy
spacing between the two adjacent levels, and the lowest spin excitation
possesses an effective spin 1/2, thus inducing a Kondo-like ZBA in the
differential conductance.

Similar effect is possible in vertical quantum dots for which the singlet
and triplet states may be close in energy both at even and odd occupation.
The influence of an external magnetic field on the orbital part of the wave
functions of electrons in vertical quantum dots is, in general, 
more pronounced than the
Zeeman effect. Hence, singlet-triplet level crossing 
are induced by this field, causing the emergence of
 Kondo scattering at even filling or its
enhancement at odd filling \cite{Sas00}. The theory of Kondo tunneling
through vertical quantum dots in external magnetic field is developed in
Refs \onlinecite{Gita00,Eto00,Pust00}.

In the present paper we explore yet another device which 
manifests the Kondo effect in QD with even electron number $N$, 
namely, a QD with two wells which is refereed to as  double
quantum dot (DQD). 
A systematic treatment of the physics 
of DQD with even N coupled to metallic leads is presented below. 
Special attention
is given to the symmetry properties of DQD
and its representation as a quantum spin
rotator. It is well known \cite{Glazr88b} that the tunneling Hamiltonian for
a QD can be mapped on the Kondo Hamiltonian in the O-diamond window of the
QD. In the E-diamond window, the same procedure of eliminating the charged
virtual states results in the four-state Hamiltonian of doubly occupied dot
where the singlet $S=0$ and triplet $S=1$ levels are intermixed by
second-order tunneling. As is shown in Ref. \onlinecite{KA01}, this
effective Hamiltonian possesses the dynamical symmetry $SO(4)$ of a spin
rotator. As a Kondo scatterer, spin rotator possesses new properties in
comparison with localized spins obeying SU(2) symmetry. The magnetic field
induced Kondo effect mentioned above is a manifestation of a "hidden
symmetry" which is a footprint of the $SO(4)$ group.

In Ref. \onlinecite{KA01} a special case was considered, namely, an
asymmetric DQD formed by two dots of different radii in a parallel geometry
coupled by tunneling interaction with even occupation $N=\nu_l+\nu_r$ ($l,r$
stands for left and right respectively). Moreover, it has been assumed that
the strong Coulomb blockade exists in one dot whereas tunneling contact with
the metallic leads exists in the other dot. Here we will address more
general situations and compare several representations in terms of effective
spin Hamiltonians. It will be shown that unusual ZBA can arise in generic
DQD structures. In particular, the Kondo effect induced by quantum dots with 
$SO(4)$ spin rotational properties exists in asymmetric DQD also when both
the $l$ and $r$ dots are coupled with the leads and the Coulomb blockade is
strong enough in both of them. The main precondition for the emergence of
the Kondo effect in this case is the sizable difference in ionization
energies of the two dots. This quantity can, in fact, be tuned by an
application of a suitable gate voltage to one of them. The same effect can
be achieved also in a symmetric DQD with even occupation in a parallel
geometry provided the axial symmetry of the system is broken by the
difference in gate voltages applied to the right and left dot ($V_g^{r,l}$,
respectively).

DQD oriented parallel to the lead surfaces were fabricated several years ago 
\cite{Hoff95,Molen95}. Two main resonance effects were noticed in such
electric circuit. First, one of the dots (say, right) can be used as an
electrometer \cite{Molen95}. Scanning $V_{g}^{r}$ at fixed $V_{g}^{l}$,
Coulomb oscillations can be induced both in the right and left dot because
the interdot capacitive coupling changes the positions of Coulomb resonance
in both of them. As a result, the step-wise structure of the conductance
acquires more complicated form. The Coulomb blockade windows between the
resonances in the Coulomb energy of the dot ${\cal E}_{\nu _{r},\nu
_{l}}(V_{g}^{r},V_{g}^{l})$ form an ''egg-carton'' pattern \cite{Hoff95}
where the vertices connect the windows with charge configurations $(\nu
_{r},\nu _{l}),(\nu _{r},\nu _{l}-1),(\nu _{r}+1,\nu _{l}-1)$. The lines $%
{\cal E}_{\nu _{l},\nu _{r}}\approx {\cal E}_{\nu _{l}+1\nu _{r},}$ are the
regions where the Coulomb resonance induced by $V_{g}^{r}$ allows tunneling
through the left dot. Second, the resonance ${\cal E}_{\nu _{r},\nu
_{l}}\approx {\cal E}_{\nu _{r}+1,\nu _{l}-1}$ allows cotunneling through
the right and left dots, which is the precondition for the Kondo effect due
to appearance of pseudospin-like configuration of the DQD \cite{WW00}. Then,
manipulating with $V_{g}^{r}$, one can induce the third transition $\nu
_{l}-1,\nu _{r}+1\rightarrow \nu _{l},\nu _{r}$ thus closing the loop and
organizing the ''electron pump'' which transfer single electron from one dot
to another (see \cite{Flens99} and references therein).

The picture becomes even richer if the tunneling between the right and left
wells of the DQD is taken into account. Then the dot can be treated as an
artificial molecule where the interdot tunneling results in formation of
complicated manifold of bonding and anti-bonding states \cite{Part00} which
modifies its charge degrees of freedom. Besides, it induces an indirect
exchange, thus modifying the Kondo resonances when the dots are placed in
series \cite{Geim99}. It will be shown below that the interdot tunneling in 
{\it parallel} geometry results in the appearance of Kondo precursor of
Coulomb resonance along the lines ${\cal E}_{\nu_r,\nu_l}\approx {\cal E}%
_{\nu_r+1,\nu_l-1}$ provided there exists direct tunneling coupling $V$
between the left and right dot. We consider the simplest case of $%
\nu_l=\nu_r=1$ in a neutral ground state of DQD. It will be shown that
unconventional Kondo resonance occurs under condition $V/[{\cal E}_{1,2}- 
{\cal E}_{1,1}]\ll 1$. Moreover, this kind of Kondo resonance can appear
also in the middle of the Coulomb window for the right dot, provided the
capacitance of the left dot exceeds essentially that of the right dot\cite
{blik}. In both cases, the DQD possesses the symmetry of a spin rotator.

In section 2 the various setups of DQD are introduced and the Hamiltonian
describing the DQD is written down within the framework of a generalized
Anderson model. The phase diagram of charging states in the left and right
gate voltages plan is schematically drown, and the regions of Kondo
resonance are indicated. In the first part of Section 3 the spectrum of the
isolated dot with even occupation is discussed. The second part is devoted
to the derivation and solution of renormalization group (RG) equations for
the DQD. The central result of this sub section is a demonstration of
possible singlet triplet level crossing due to tunneling. When the
renormalized energies are below the reduced band edge, renormalization
stops, and charge fluctuations are suppressed. This is the Schrieffer-Wolff
regime, and a derivation of an effective spin Hamiltonian is executed in
section 4. In the first part, the spin Hamiltonian is given in terms of two
vector operators and is shown to have the $SO(4)$ symmetry of a spin
rotator. It is followed by a short subsection in which the renormalization
group flow of coupling constants is explained and the Kondo temperature is
derived. Then, in the third subsection, a two spin representation is
suggested, in which the occurrence of two spin 1/2 operators just reflects
the fact that the algebra $o_{4}$ is a direct sum of two $o_{3}$ algebras.
In the fourth subsection, the possibility of arriving at the Kondo effect in
finite magnetic field is discussed, leading to a third representation of the
spin Hamiltonian. The question of whether a DQD with two electrons can be
regarded as a real two site Kondo system (even if the DQD is highly
asymmetric) is discussed in section 5. In particular, a stringent comparison
is made with the two spin representation mentioned in section 4. The paper
is concluded in section 6. Some technical details of various calculations
are relegated to the Appendix.

\newpage
\section{Models of double quantum dots with singlet ground state}

Two models considered in this work are sketched in Fig.1. We will refer to a
system (a) with zero gate voltages as a ''symmetric'' DQD. The same system
with finite but unequal gate voltages $V_{g}^{l,r}$ will be called a
''biased'' DQD and the pair of dots with different radii shown in Fig. 1b
will be referred to as ''asymmetric'' DQD. 
\begin{figure}[htb]
\centering
\includegraphics[width=60mm,
height=60mm,angle=0,
]{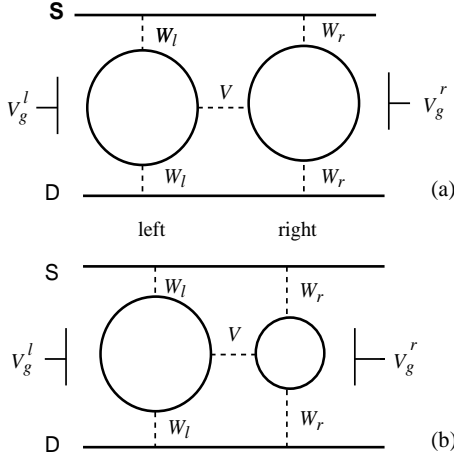}
\narrowtext
\caption{Double quantum dots in parallel geometry. 
Left $(l)$ and right $(r)$ dots
are coupled by tunneling $V$ to each other and by tunneling $W_{l,r}$ to the
source (S) and drain (D) electrodes. $V_g^{l,r}$ are the gate voltages.
(a) symmetric dot, (b) asymmetric dot.}
\label{dqd1}
\end{figure}
In all cases the DQD is described
by a generalized Anderson tunneling Hamiltonian which takes into account the
internal structure of the DQD, 
\begin{equation}
H=H_{b}+H_{t}+H_{d}+H_{g}.  \label{1.1}
\end{equation}
The first term, $H_{b}$ is related to the lead electrons, 
\begin{equation}
H_{b}=\sum_{k\sigma ,\alpha }\varepsilon _{k\sigma ,\alpha }c_{k\sigma
,\alpha }^{\dagger }c_{k\sigma \alpha }~,  \label{1.1a}
\end{equation}
where $\alpha =s,d$ denotes electrons from the source and drain electrodes 
\cite{foot} and $\sigma =\pm $ 
is the spin index. The second term, $H_{t}$ is
the tunneling Hamiltonian, 
\begin{equation}
H_{t}=\sum_{i=l,r}\sum_{k\sigma }\left( W_{ki}c_{k\sigma }^{\dagger
}d_{i\sigma }+h.c.\right) ~.  \label{1.1b}
\end{equation}
Here $c_{k\sigma }=2^{-1/2}(c_{k\sigma ,s}+c_{k\sigma ,d})$, and~ $%
W_{ki}=W_{k\alpha ,i}/(W_{ks,i}^{2}+W_{kd,i}^{2})^{1/2}.$ The third term, $%
H_{d},$ describes the isolated DQD. In the present context, the quantum dot
is a ''molecule'', containing $N=\nu _{l0}+\nu _{r0}$ electrons in a
neutral ground state. The capacitive interaction {\em between the two 
wells of the} DQD
is assumed to be strong enough to
suppress the fluctuations of electron tunneling induced occupation in the
windows between the Coulomb resonances of tunneling amplitude. We consider
DQD with even $N$, so that, generically, the ground state of an isolated DQD
is a spin singlet. The isolated dot is then described by the Hamiltonian, 
\begin{equation}
H_{d}=\sum_{i=l,r}\sum_{\sigma }\epsilon _{i}n_{i\sigma }+V\sum_{i\neq
j}d_{i\sigma }^{\dagger }d_{j\sigma }+H_{corr},  \label{1.2}
\end{equation}
in which $V$ is the inter-well constant tunneling amplitude. The 
capacitive interaction
within the DQD is described by the term 
\begin{equation}
H_{corr}=\frac{1}{2}\sum_{i}Q_{i}n_{i}(n_{i}-1)+Q_{lr}\delta n_{l}\delta
n_{r}.  \label{1.2a}
\end{equation}
Here $n_{i}=\sum_{\sigma }d_{i\sigma }^{\dagger }d_{i\sigma }$, and $\delta
n_{i}=n_{i}-\nu _{i0}$ is the deviation of electron distribution from the
neutral charge configuration $\nu _{i0}$ for a given DQD. Moreover, $%
Q_{i}=e^{2}/2C_{i}$ is the charging energy of the dot $i$ whose capacitance
is $C_{i},$ and $Q_{lr}$ is the capacitive coupling between the left and
right dots. 
The simplest configuration which contains in a nutshell all the
complicated physics of many-body interactions arising in a course of
tunneling is $N=2$, $\nu _{l0}=\nu _{r0}=1$. 
This case, for which  $\delta n_{i}=n_{i}-1.$, will be given a special
attention below. Finally, the term $H_{g}$
represents the gate voltage energy. We consider symmetric and asymmetric
DQDs formed by wells of equal and different radii respectively (Fig. 1).
Hence, generically, the gate potential $H_{g}$ is asymmetric, 
\begin{equation}
H_{g}=\sum_{i}V_{g}^{i}N_{i},~~~V_{g}^{l}\neq V_{g}^{r}~.  \label{1.2b}
\end{equation}
It is, in fact, useful to include the gate potential (\ref{1.2b}) in the
position of the one-electron energy levels, $\varepsilon _{i}=\epsilon
_{i}+V_{g}^{i}$. Then, by tuning the gate voltage one can change the energy
difference $\Delta =\varepsilon _{l}-\varepsilon _{r}$ or, in other words,
redistribute the electron density between the left and right wells of the
DQD.

It is assumed that in equilibrium and at zero gate voltages, each dot is
filled by one electron and the Fermi level of the leads is in the middle of
the Coulomb blockade window. The energy levels of a symmetric DQD with
uncoupled dots $(Q_l=Q_r=Q,~V=0)$) are shown in the upper panel (a) of Fig.
2. These levels may be shifted relative to each other and to the Fermi level 
$\varepsilon_F$, and each level crossing $\varepsilon_i-\varepsilon_F$
corresponds to recharging of the dot $i$. If electron exchange between the
right dot and the leads is blocked \cite{Hoff95}, the charge transfer
resonance between the states $\{1,1\}$ and $\{0,2\}$ occurs when $%
\varepsilon_l=\varepsilon_r+Q$ (see also \cite{Uga99}). In the general case
(Fig.1a), additional electron appears in the dot $i$ when the levels $%
\varepsilon_i+Q$ and $\varepsilon_F$ cross.
\begin{figure}[htb]
\centering
\includegraphics[width=75mm,
height=60mm,angle=0,
]{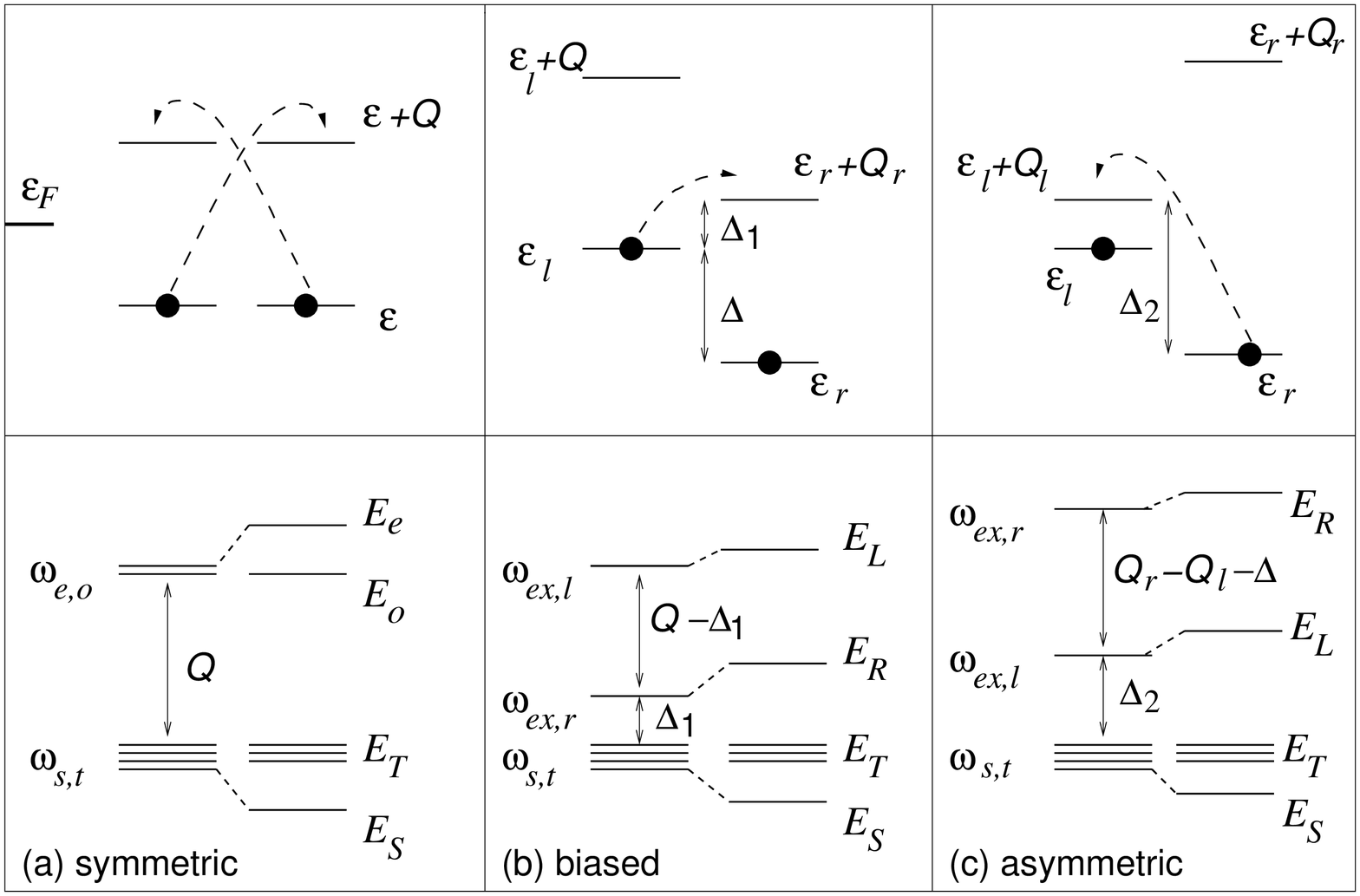}
\caption{Energy level scheme for symmetric $(a)$, biased $(b)$ and asymmetric
$(c)$ DQD. Upper panel: filled and empty one-electron levels. Dashed arrows
indicate charge-transfer excitons. Lower panel: two-electron states of isolated
and coupled left and right dots.}
\label{dqd2}
\end{figure}
In the absence of interdot tunneling, $V=0$, one can easily obtain the
effective spin Hamiltonian for the DQD with $N=2$ in the ground state far
deep in the Coulomb blockade windows. This is the two-site Kondo Hamiltonian
in the window $\{1,1\}$ and the single site Kondo Hamiltonians in the
windows $\{2,1\}$ and $\{1,2\}$. In the latter case of charged DQD occupied
by odd number of electrons, tunneling through the left (right) dot is
blocked, but a Kondo-type resonance compensates for the Coulomb blockade and
opens a tunneling channel through the right (left) dot. In the former case
of neutral DQD with even occupation the possibility of Kondo tunneling is
determined by the relative strength of the on-site indirect exchange $J_i$
between the spins $S_i$ of singly occupied dots and conduction electrons in
the reservoir on the one hand, and the sign and magnitude of the intersite
RKKY exchange $J_{lr}$ on the other hand \cite{Varma}. Both these parameters
are predetermined by the tunnel coupling constants $W_{ki}$ with the band
electrons in the reservoir, but one can modify them by varying the gate
voltages and interdot distance.

The interdot coupling significantly modifies this picture. It favors the
singlet spin state in the middle of the Coulomb blockade window $\{1,1\}$
thus eliminating the Kondo tunneling at zero gate voltages. At finite $%
|V_g^l - V_g^r|$ the values of $\nu_i$ deviate from the integer values near
the boundaries between the different charge sectors. Increasing negative
gate voltages $V_g^l$ or $V_g^r$, one can bias the charge distribution in
favor of left or right dot, respectively, without changing the total number
of electrons. As a result, with increasing $|V_g^l - V_g^r|$ one reaches the
region of states with small charge transfer gap $\Delta_1\equiv Q+
\varepsilon_r-\varepsilon_l \ll Q$. The energy levels of such "biased" DQD
are shown in the upper panel of Fig. 2b. These states occupy the upper right
corner of the window $\{1,1\}$ hatched in fig.3a. Here the zero energy
configuration illustrated by Fig. 2a corresponds to the coordinate origin.
The virtual charge transfer excitations (dashed arrow in the upper panel of
Fig.2b) significantly influence the tunneling through the DQD. It will be
shown below that a novel type of Kondo resonance arises in this area of the
sector $\{1,1\}$. The "biased" DQD in this sector behaves like a spin
singlet at high temperatures and excitation energies, and demonstrates the
properties of spin one triplet partially screened by the Kondo tunneling at
low energies and temperatures $T<T_K$. The Kondo temperature $T_K$ is a
function of $V_{g}^i$. The Kondo "isotherm" $T_K(V_g^l,V_g^r)$ is presented
by the dashed line in Fig. 3a.
\begin{figure}[htb]
\centering
\includegraphics[
height=80mm, width=60mm,angle=0,
]{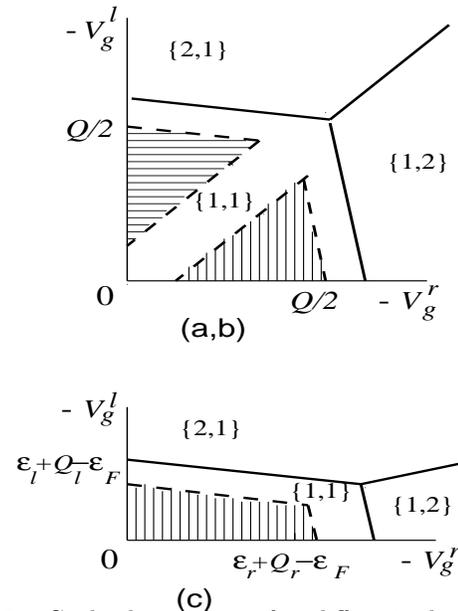}
\caption{Coulomb windows for different charge states $\{\nu_l, \nu_r \}$ of
symmetric (a,b) and asymmetric (c) DQD. Hatched regions indicate the domains where
the Kondo effect exists.}
\label{dqd3}
\end{figure}
Similar effect exists for the asymmetric DQD (Fig. 1c) where the two coupled
dots have different radii $r_l\gg r_r$ and hence different blockade
energies, $Q_l \ll Q_r$. The energy levels of an isolated DQD are shown in
the upper panel of Fig. 2c, and the corresponding recharging map is
presented in Fig. 3b. Here the hatched area also marks the region of the map
where Kondo effect arises in spite of the even number of electrons in the
dot. The Kondo isotherms in this case are parallel to the Coulomb resonance
line.

Conventional approach for the description of the Kondo effect in a two-site
quantum dot starts with the {\it two-center} Hamiltonian 
\begin{equation}
H_{d0}=H_l + H_r,  \label{3.5a}
\end{equation}
and treats $H_t$ in terms of a two channel tunneling operator, 
\begin{equation}
H_{t}=H_{tl} + H_{tr}.  \label{3.5b}
\end{equation}
The interdot interaction $H_{lr}$ is considered as a coupling between two
resonant Anderson centers. If the left and right dot each contains an odd
number of electrons (as in our simplified model with $\nu_{l,r}=1$), the
Kondo tunneling is possible through each dot separately. The exchange part
of interdot coupling maps our Hamiltonian onto the two-site Kondo model.
This coupling can be both of ferromagnetic and antiferromagnetic type. In
the latter case the interplay between $H_t$ and $H_{lr}$ results in
suppression of Kondo tunneling through the left and/or right well of the
DQD. The phase diagram of the two-site Kondo model is discussed in numerous
papers \cite{Varma}.

In the model discussed here, the interdot interaction is represented by the
term $H_{lr}=V\sum_{i\neq j}d_{i\sigma }^{\dagger }d_{j\sigma }$ till we
remain in the charge sectors $\{1,1\},\{1,2\},\{2,1\}$ of Fig. 3. It is
obvious that this coupling suppresses Kondo tunneling through the symmetric
DQD at zero gate voltages (point ''0'' in Fig. 3a) because the effective
indirect exchange interaction which arises due to virtual excitations of
charged states $\{0,2\},\{2,0\}$ is of antiferromagnetic sign, $%
J_{lr}=2V^{2}/Q$ like in the Heitler-London limit for a hydrogen molecule or
in the half-filled Hubbard model. As a result, the ground state of a DQD is
a spin singlet, and the gap $\delta =E_{T}-E_{S}=J_{lr}$ which divides the
triplet excitation from the singlet spin ground state prevents the formation
of a Kondo resonance. It will be demonstrated below that this is not so in
the case of {\it strongly asymmetric} DQD (hatched regions in Fig. 3a,b),
where the crossover to a triplet state is induced by the tunneling $H_{t}$.

To describe this crossover, it is more convenient first to diagonalize the
dot Hamiltonian $H_{d}$, i.e. to express it in the form 
\begin{equation}
H_{d}=\sum_{N,\Lambda }E_{N\Lambda }|N\Lambda \rangle \langle N\Lambda |,
\label{1.01}
\end{equation}
recalling that
$N$ is the number of electrons in a given charge state of 
the DQD whereas $%
\Lambda $ stands for a set of quantum numbers 
which characterize the many-electron
configuration $d_{l}^{\nu _{l}}d_{r}^{\nu _{r}}$ in the presence of interdot
coupling. In order to get compact form for some equations, we introduce
Hubbard projection and configuration change operators 
\begin{equation}
X^{N\Lambda ,N^{\prime }\Lambda ^{\prime }}=|N\Lambda \rangle \langle
N^{\prime }\Lambda ^{\prime }|~.  \label{1.010}
\end{equation}
The diagonal terms $X^{N\Lambda ,N\Lambda }$ are conventional projection
operators, while the off-diagonal operators change electron configuration of
the dot. The tunneling term $H_{t}$ (\ref{1.1b}) can now be rewritten in the
form, 
\begin{equation}
H_{t}=\sum_{N,\Lambda }\sum_{N^{\prime },\Lambda ^{\prime }}\sum_{k\sigma
}\left( W_{k\sigma }^{N\Lambda ,N^{\prime }\Lambda ^{\prime }}X^{N\Lambda
,N^{\prime }\Lambda ^{\prime }}c_{k\sigma }+h.c.\right) ~.  \label{1.03}
\end{equation}
The matrix elements $W_{k\sigma }^{N\Lambda ,N^{\prime }\Lambda ^{\prime }}$
are nonzero for states in adjacent charge sectors of the eigen space of $%
H_{d}$, so that $N=N^{\prime }+1$. In this approach, the DQD is treated as a
''resonance impurity'' in the framework of the conventional Anderson model,
and its specific features are manifest in a characteristic energy spectrum $%
E_{N\Lambda }$ which includes contributions due to interdot tunneling and
Coulomb blockade.

\section{Doubly occupied DQD as an Anderson impurity}

\subsection{Energy levels and wave functions}

Let us now employ the above approach to the doubly occupied DQD with $N=2$
in a charge sector $\{1,1\}$ of the Coulomb blockade diagram (Fig. 3). The
dot Hamiltonian (\ref{1.2}),(\ref{1.2a}),(\ref{1.2b}) can be exactly
diagonalized by using the basis of two-electron wave functions 
\begin{eqnarray}
|s\rangle &=&\frac{1}{\sqrt{2}}\sum_{\sigma }\sigma d_{l\sigma
}^{\dagger}d_{r\bar\sigma }^{\dagger }|0\rangle ~;  \label{1.02} \\
|t_{0}\rangle &=&\frac{1}{\sqrt{2}} \sum_{\sigma }d_{l\sigma
}^{\dagger}d_{r\bar\sigma }^{\dagger }|0\rangle, \quad |t_{\sigma }\rangle
=d_{l\sigma}^{\dagger } d_{r\sigma }^{\dagger }|0\rangle ~;  \nonumber \\
|ex_l\rangle & = & d_{l\uparrow }^{\dagger }d_{l_\downarrow }^{\dagger }
|0\rangle, \quad |ex_r\rangle = d_{r\uparrow }^{\dagger }d_{r_\downarrow
}^{\dagger }|0\rangle ~.  \nonumber
\end{eqnarray}

Generically, the spectrum of a {\it neutral} DQD consists of a singlet
ground state $|S\rangle $, a low-energy spin-one triplet exciton $|T\mu
\rangle $ and two high-energy charge-transfer singlet excitons $%
|Ex_{l}\rangle $ and $|Ex_{r}\rangle $. The corresponding two-electron wave
functions are the following combinations: 
\begin{eqnarray}
|S\rangle  &=&a_{ss}|s\rangle +a_{sl}|ex_{l}\rangle +a_{sr}|ex_{r}\rangle ~;
\label{1.3} \\
|T0\rangle  &=&|t_{0}\rangle ,~~|T\pm \rangle =|t_{\pm }\rangle ~;  \nonumber
\\
|Ex_{l}\rangle  &=&a_{ll}|ex_{l}\rangle +a_{ls}|s\rangle ,~|Ex_{r}\rangle
=a_{rr}|ex_{l}\rangle +a_{rs}|s\rangle .  \nonumber
\end{eqnarray}
In the special case of symmetric DQD ($\varepsilon _{l}=\varepsilon
_{r},~Q_{l}=Q_{r}$), the axial symmetry allows one to introduce even $(e)$
and odd $(o)$ excitonic states 
\[
|ex_{e,o}\rangle =\frac{1}{\sqrt{2}}(|ex_{l}\rangle \pm |ex_{r}\rangle ).
\]
The interdot tunneling leaves intact odd singlet state $|ex_{o}\rangle $ as
well as odd triplet states $|T\mu \rangle $. As a result, one has instead of
(\ref{1.3}), 
\begin{eqnarray}
|S\rangle  &=&a_{ss}|s\rangle +a_{se}|ex_{e}\rangle ,~~~|T\mu \rangle
=|t_{\mu }\rangle ,  \nonumber \\
|Ex_{e}\rangle  &=&a_{ee}|ex_{e}\rangle +a_{es}|s\rangle ,~~~|Ex_{o}\rangle
=|ex_{o}\rangle ,  \label{1.4}
\end{eqnarray}
(see Fig. 2a). We are mainly interested in the limiting cases of strongly
biased symmetric DQD where the interdot tunneling results in sizable
charge transfer between the left and right dot with charge transfer energy $%
\Delta _{1}=\varepsilon _{r}+Q-\varepsilon _{l}$ (Fig. 2b) and asymmetric
dot with charge transfer energy $\Delta _{2}=\varepsilon
_{l}+Q_{l}-\varepsilon _{r}$ (Fig. 2c). The virtual charge transfer
transitions which contribute to the lowest part of the energy spectrum are
marked by the dashed arrows. Charge fluctuations are negligible when 
\begin{equation}
\beta =V/Q\ll 1,~~~\beta _{1}=V/\Delta _{1}\ll 1,~~~\beta _{2}=V/\Delta
_{2}\ll 1,  \label{2.1}
\end{equation}
in cases (a), (b), (c) respectively \cite{transf}. The expansion
coefficients $a_{ij}$ in this limit are calculated for all three cases in
the Appendix (eqs. \ref{A.6},\ref{A.8},\ref{A.10},  see also comment and
references \cite{Ivan}). In a symmetric configuration (a) the two-electron
levels which correspond to the bare states of a symmetric DQD form a
low-energy quartet $\omega _{s,t}=2\varepsilon $, $\omega
_{ex_{e,o}}=2\varepsilon +Q$. The odd states remain unrenormalized as a
result of interdot tunneling, whereas the even states undergo a level
repulsion. In the limit of small $\beta =V/Q\ll 1$ 
\begin{eqnarray}
E_{S} &=&2\varepsilon -2V\beta ,~~~E_{T}=2\varepsilon ,  \label{2.2} \\
E_{o} &=&2\varepsilon +Q,~~~E_{e}=2\varepsilon +Q+2V\beta ,  \nonumber
\end{eqnarray}
(see lower panel of Fig. 2a). In the case (b) of biased DQD, the
two-electron bare energy levels are arranged as is shown in the lower panel
of Fig. 2b: 
\[
\omega _{s,t}=\varepsilon _{l}+\varepsilon _{r},~\omega _{ex,r}=\varepsilon
_{r}+Q,~\omega _{ex,l}=\varepsilon _{l}+Q~.
\]
The parity is broken in this case, and the singlet state $|s\rangle $ is now
hybridized with both excitonic states. In the limit (\ref{2.1}) one has, 
\begin{eqnarray}
E_{S} &=&\varepsilon _{l}+\varepsilon _{r}-2\beta _{1}V,~~~E_{T}=\varepsilon
_{l}+\varepsilon _{r},  \label{2.3} \\
E_{R} &=&2\varepsilon _{r}+Q+2\beta _{1}V,~~~E_{L}=2\varepsilon
_{l}+Q+2\beta _{1}^{\prime }V~.  \nonumber
\end{eqnarray}
The role of the ''left'' exciton $|E_{L}\rangle $ in the low-energy
processes is negligibly small. In the same spirit, all terms $\sim \beta
_{1}^{\prime }=V/(\varepsilon _{l}-\varepsilon _{r}+Q)$ will be neglected in
pertinent calculations below.

The general scheme of energy eigenvalues in case (c) of an asymmetric DQD is
similar to that of case (b), but here, the first singlet excitation state is
a charge-transfer exciton in the left dot. So the bare two-electron spectrum
is $\omega _{s,t}=\varepsilon _{l}+\varepsilon _{r}$, $\omega
_{ex,l}=2\varepsilon _{l}+Q_{l}$, $\omega _{ex,r}=2\varepsilon _{r}+Q_{r}$,
and the hybridized states are approximately given by, 
\begin{eqnarray}
E_{S} &=&\varepsilon _{l}+\varepsilon _{r}-2\beta _{2}V,~~~E_{T}=\varepsilon
_{l}+\varepsilon _{r},  \label{2.4} \\
E_{L} &=&2\varepsilon _{l}+Q_{l}+2\beta _{2}V,~~~E_{R}=2\varepsilon
_{r}+Q_{r}+2\beta _{2}^{\prime }V,  \nonumber
\end{eqnarray}
(see lower panel of Fig. 2c). In this case we neglect the contribution of
''right'' exciton $|E_{R}\rangle $ and all terms $\sim \beta _{2}^{\prime
}=V/(\varepsilon _{r}-\varepsilon _{l}+Q_{r}).$ 

It is seen from Eqs (\ref{2.2}), (\ref{2.3}), (\ref{2.4}) for the energy
spectrum of an isolated DQD that the low-energy excitations with energy $%
\delta _{s}=E_{T}-E_{S}$ are dominantly of spin character, whereas the
charge excitations $E_{e,o}$ in case (a) and $E_{L,R}$ in cases (b,c) are
separated from the ground state by the gaps $\delta _{ch}\gg \delta _{s}$ in
all three cases under consideration (lower panels of Fig. 2a-c). This same
kind of ``spin-charge separation'' persists when the DQD is hybridized (via $%
H_{t}$) with itinerant electrons in the metallic reservoirs, on which we now
focus our attention.

\subsection{Renormalization of energy levels}

The spectrum of electrons in the reservoirs is continuous and form a band
with bandwidth $2D_{0}$. In accordance with the renormalization group (RG)
procedure widely used in the conventional Anderson model, the low energy
physics can be exposed by integrating out the high-energy charge excitations
in a framework of poor man's scaling technique \cite{Hald78}. This procedure
implies renormalization of the energy levels and coupling constants of the
Hamiltonian (\ref{1.1}) by mapping the initial energy spectrum $-D_0<
\varepsilon < D_0$ onto a reduced energy band $-D_0+|\delta D| <\varepsilon
< D_0-|\delta D|$.

The mapping procedure results in the following equations for the singlet and
triplet renormalized energies of the DQD: 
\begin{equation}
E_{\Lambda }\approx E_{\Lambda }^{(0)}+\sum_{\lambda }\sum_{q\sigma }\frac{%
|W_{q\sigma }^{\Lambda \lambda }|^{2}}{E_{\Lambda }-\epsilon _{q}-E_{\lambda
}},  \label{3.1a}
\end{equation}
where $E_{\Lambda }^{(0)}$ is the energy before renormalization, $%
q=q_{u},q_{b}$ are electron momenta such that $\epsilon _{q}$ belong to the
layers $|\delta D|$ near the top or the bottom of the conduction band
respectively. They appear as intermediate virtual states in the processes of
positive and negative ionization of the DQD. The index $(\Lambda =S,T\mu )$
in these equations is reserved for the neutral two-electron states (\ref{1.3}%
) of the DQD, whereas the positively and negatively charged states with one
and three electrons are designated by the index $\lambda $. The wave
functions and energy levels of these states as well as the matrix elements $%
W_{q,\sigma }^{\Lambda \lambda }$ are calculated in the Appendix. Figure 4
illustrates the processes involved in the level renormalization in all three
cases under consideration. Note that these RG equations are uncoupled in
this order. In accordance with the poor man's scaling approach \cite{Hald78}
only the virtual transitions with energy $\sim D$ are relevant, and the
estimate of the sum in the r.h.s. of eq. (\ref{3.1a}) gives 
\begin{equation}
E_{\Lambda }=E_{\Lambda }^{(0)}-\frac{\Gamma ^{\Lambda }|\delta D|}{D},
\label{3.1b}
\end{equation}
where $\Gamma ^{\Lambda }=\pi \rho _{0}|W^{\Lambda }|^{2}$, $\rho _{0}$ is
the density of electron states in the reservoir, which is taken to be
constant, and $W^{\Lambda }$ are effective tunneling matrix elements
calculated in the Appendix.

The crucial difference between the symmetric configuration (a) and the
asymmetric configurations (b,c) is that the tunneling amplitudes of the
processes involved in renormalization (3.1a) are different for singlet and
triplet states of the DQD. In the symmetric case (a), the left and right dot
states are involved in renormalization of the two-electron states on an
equal footing. The relevant processes are $|S\rangle \to |eq_u,1e\rangle$, $%
|S\rangle \to |oq_u,1o\rangle$, $|T\rangle \to |eq_u,1e\rangle$, $|S\to
eq_u,1e\rangle$. The one-electron tunneling transitions that give dominant
contribution to these processes are shown by the dashed arrows in Fig. 4a.
\begin{figure}[htb]
\centering
\includegraphics[width=60mm,
height=80mm,angle=0,
]{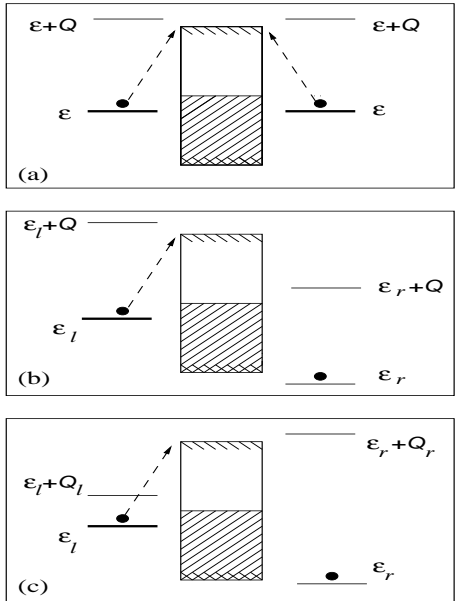}
\caption{The particle states, which are removed from half-filled
conduction band on reducing the bandwidth by $|\delta D|$. The one-electron
levels renormalized as a result of this process are shown by bold lines.
(a) symmetric DQD, (b) biased DQD, (c) asymmetric DQD.}
\label{dqd4}
\end{figure}
As a result, the tunneling rate in this case is 
\begin{equation}
\Gamma_S=\Gamma_T\approx \pi\rho_0(|W_l|^2+|W_r|^2).  \label{3.7a}
\end{equation}

In the asymmetric configurations (b) and (c) the even-odd symmetry is
broken, and the Coulomb blockade in one center controls the tunneling
through the other one \cite{Hoff95,Molen95}. Processes with energy $\sim D$
involve only the electrons from the left dot. In case (b) the relevant
processes are $|S\rangle \to |q_{u}\sigma,1{b\bar\sigma}\rangle,$ $%
|T0\rangle \to |q_{u}\sigma,1{b\bar\sigma}\rangle,$ $|T\pm\rangle \to
|q_{u}\pm,1_{b}\pm\rangle,$ and the tunneling transitions, which give the
dominant contribution to these processes are shown by the dashed arrow in
Fig. 2b. The same kind of asymmetry takes place in case (c) (dashed arrows
in Fig. 4c). As a result, one has, instead of (\ref{3.7a}), 
\begin{equation}
\Gamma_T\approx \pi\rho_0 |W_l|^2 \Gamma_S=a_{ss}^2\Gamma_T.  \label{3.7b}
\end{equation}
Here the coefficient $a_{ss}<1$ is a measure of charge transfer from the
left dot to the right dot due to admixture of singlet excitonic states to
the ground state singlet [see Eqs. (\ref{3.4}), (\ref{3.7}) in the Appendix].
Iterating the renormalization procedure (\ref{3.1b}), one comes to the
couple of differential scaling equations 
\begin{equation}
\frac{dE_\Lambda}{d\ln D}=\frac{\Gamma_\Lambda}{\pi},~~~\Lambda=T,S~,
\label{3.6}
\end{equation}
which describe the evolution of the two-electron energy states with reducing
the energy scale of the band continuum. These equations describe not only
the renormalization of the low-energy two-electron spin states but also the
change of the one-electron transition energies $E_\Lambda-E_\lambda$,
because the one-electron states E$_{\lambda=1b} =\varepsilon_r-O(\beta)$ are
deep under the Fermi level, and the reduction of the energy scale does not
influence them.

Scaling equations of the type (\ref{3.6}) were analyzed in Ref. %
\onlinecite{KA01} for a specific case of "Fulde molecule" or double shell
quantum dot (DSD), where the electrons in one shell are subject to strong
correlation effect (Coulomb blockade) whereas the loosely bound electrons in
the second shell are responsible for tunneling, and tunneling to the leads
is allowed only for the second shell. The model (c) is a natural extension
of DSD because in the lowest approximation in the interdot interaction the
tunneling through the right dot gives no contribution to level
renormalization. In case (b), both left and right dot contribute to the
renormalization procedure, but the crucial property of scaling equations, $%
\Gamma_S < \Gamma_T $ (see eq. \ref{3.7b}) is shared by both configurations.

The scaling invariants for equations (\ref{3.6}) are 
\begin{equation}
E_{\Lambda }^{\ast }=E_{\Lambda }(D)-\frac{\Gamma _{\Lambda }}{\pi }\ln
\left( \frac{\pi D}{\Gamma _{\Lambda }}\right) .  \label{3.6b}
\end{equation}
Here the scaling constants have to be chosen to satisfy the boundary
condition $E_{\Lambda }(D_{0})=E_{\Lambda }^{(0)}$. Due to relation (\ref
{3.7b}) the energy $E_{T}(D)$ decreases with $D$ faster than $E_{S}(D)$, so
that the two scaling trajectories $E_{\Lambda }$ cross at a a certain
bandwidth $D=D_{c}$ estimated as 
\begin{equation}
\frac{\Gamma _{T}-\Gamma _{S}}{\pi }\ln \frac{D_{0}}{D_{c}}%
=E_{T}^{(0)}-E_{S}^{(0)}\equiv \delta _{0}.  \label{3.6a}
\end{equation}
According to calculations performed in Ref. \onlinecite{KA01}, this level
crossing can occur either before or after the crossover to the
Schrieffer-Wolff regime when the one-electron energies $E_{\Lambda
}(D)-E_{1b}$ exceed the half-width of the reduced continuous spectrum $\bar{D%
}\sim |E_{\Lambda }(\bar{D})-E_{1b}|$. In both cases, the charge degrees of
freedom are quenched for excitation energies within the interval $-\bar{D}%
<\varepsilon <\bar{D},$ and Haldane's renormalization procedure should be
replaced by the Anderson poor man's scaling \cite{Anders70}.

\section{Spin Hamiltonian for DQD}

\subsection{The quantum rotator representation}

The Schrieffer-Wolff transformation \cite{SW69} for the configuration of two
electron states of a DQD projects out those states of the dot having one or
three electrons and maps the Hamiltonian $H$ onto an effective spin
Hamiltonian $\widetilde{H}$ acting in a subspace of two-electron
configurations $\langle\Lambda|\ldots |\Lambda^{\prime}\rangle$, 
\begin{equation}
\widetilde{H}= e^{{\cal S}}H e^{{\cal -S}}= H + \sum_m \frac{(-1)^m}{m!} [%
{\cal S},[{\cal S}...[{\cal S},H]]...],  \label{3.8}
\end{equation}
where 
\begin{equation}
{\cal S}=\sum_{\Lambda\lambda}\sum_{\langle k\rangle\sigma} \frac{%
(W_{\sigma}^{\Lambda\lambda})^*} {\bar{E}_{\Lambda\lambda}-\epsilon_k}%
X^{\Lambda\lambda}c_{k\sigma}+h.c.  \label{3.8a}
\end{equation}
Here $\langle k\rangle$ stands for the electron or hole states secluded
within a layer $\pm \bar{D}$ around the Fermi level. $\bar{E}%
_{\Lambda\lambda}=E_\Lambda(\bar{D})-E_\lambda(\bar{D}).$ The effective
Hamiltonian with the charged states $|\lambda\rangle=|1b\sigma\rangle,
|3b\sigma\rangle$ frozen out can be obtained within first order in ${\cal S}$%
. It has the following form, 
\begin{eqnarray}
\widetilde{H} & = & \sum_\Lambda \bar{E}_\Lambda X^{\Lambda\Lambda}
+\sum_{\langle k\rangle \sigma}\epsilon_k c^\dagger_{k\sigma}c_{k\sigma} 
\nonumber \\
& - &
\sum_{\Lambda\Lambda^{\prime}\lambda}\sum_{kk^{\prime}\sigma\sigma^{\prime}}
J^{\Lambda\Lambda^{\prime}}_{kk^{\prime}}
X^{\Lambda\Lambda^{\prime}}c^\dagger_{k\sigma}c_{k^{\prime}\sigma^{\prime}},
\label{3.9}
\end{eqnarray}
where 
\[
J^{\Lambda\Lambda^{\prime}}_{kk^{\prime}}= (W_{\sigma}^{\Lambda\lambda})^*
W_{\sigma^{\prime}}^{\Lambda^{\prime}\lambda} \left( \frac{1}{\bar{E}%
_{\Lambda\lambda}-\epsilon_k}+ \frac{1}{\bar{E}_{\Lambda^{\prime}\lambda}-%
\epsilon_{k^{\prime}}} \right). 
\]
In the charge sector $N=2$ the constraint $\sum_{\Lambda}X^{\Lambda%
\Lambda}=1 $ is valid. As is shown in \cite{Eto00,KA01}, the effective
Schrieffer-Wolff Hamiltonian of DQD describes not only the conventional
indirect exchange between localized and itinerant spins. It also contains
terms that intermix the singlet and triplet states of the quantum dot. This
mixing is due to the tunneling exchange with electrons in the metallic
reservoir.

As a result of integrating out the high-energy charge degrees of freedom,
the effective spin Hamiltonian of the DQD acquires an $SO(4)$ symmetry,
which is the dynamical symmetry of a spin rotator. Before writing down the
pertinent spin Hamiltonian a few words about a quantum spin rotator are in
order. It is known (see, e.g., \cite{Engel}) that the symmetry of a standard
quantum rotator is described by the operator of rotational angular momentum $%
{\bf L}$ and an additional vector operator ${\bf M}$. These two operators
generate the semi-simple algebra $o_{4}$, they are orthogonal, ${\bf L}\cdot 
{\bf M}=0,$ and the corresponding Casimir operator is ${\bf L}^{2}+{\bf M}%
^{2}$. The matrix elements of the operator ${\bf M}$ connect states with
different values of the orbital momentum $l\rightarrow l\pm 1$. The
existence of this second operator reflects the ''hidden angular symmetry''
of the rotator.

Similarly, the spin symmetry of the DQD is characterized not only by the
spin one vector ${\bf S}$: one can introduce a second vector operator ${\bf P%
}$ orthogonal to ${\bf S}$ which determines the matrix elements of
transitions between the different states of the rotation group $SO(4)$. In
the present case, the vector ${\bf P}=\{P_{z},P^{\pm }\}$ determines the
transitions between the singlet state and the different components of the
spin triplet. It is convenient to express the spherical components of the
vector operator ${\bf P}$ in terms of Hubbard operators $X^{\Lambda \Lambda
^{\prime }}$ (for brevity, ${\Lambda \Lambda ^{\prime }}$ will be either S
for singlet or $\mu =1,0,{\bar{1}}$ for the triplet magnetic quantum
numbers), 
\begin{eqnarray}
P^{+} &=&\sqrt{2}\left( X^{1S}-X^{S\bar{1}}\right) ,\;P^{-}=\sqrt{2}\left(
X^{S1}-X^{\bar{1}S}\right) ,  \nonumber \\
P_{z} &=&-\left( X^{0S}+X^{S0}\right) .  \label{3.9d}
\end{eqnarray}
Within the same procedure, the spherical components of the spin one operator 
${\bf S}$ are given by the following expressions, 
\begin{eqnarray}
&& S^{+}=\sqrt{2}\left( X^{10}+X^{0\bar{1}}\right) , \nonumber \\
&& S^{-}=\sqrt{2}\left(
X^{01}+X^{\bar{1}0}\right) ,~S_{z}=X^{11}-X^{\bar{1}\bar{1}}.  \label{3.9c}
\end{eqnarray}
The vector operators ${\bf P}$ and ${\bf S}$ 
obey the commutation relations of the usual $o_{4}$ Lie algebra, 
\begin{equation}
\lbrack
S_{j},S_{k}]=ie_{jkl}S_{l},\;\;[P_{j},P_{k}]=ie_{jkl}S_{l},\;%
\;[P_{j},S_{k}]=ie_{jkl}P_{l}  \label{3.9e}
\end{equation}
(here $j,k,l$ are Cartesian indices). Besides, the following relations hold,
\begin{equation}
{\bf S\cdot P}=0,\;S^{2}=2(1-X^{SS}),\;P^{2}=1+2X^{SS}.  \label{3.11}
\end{equation}
To wit, ${\bf S}$ (a spin 1) 
and ${\bf P}$ are two orthogonal vector operators in spin
space which generate the algebra $o_{4}$ in a representation specified
by the Casimir operator ${\bf S}^{2}+{\bf P}^{2}=3.$ This justifies
the qualification of DQD as a {\it spin rotator.}

Returning back to the effective spin Hamiltonian, (\ref{3.9}) it now
acquires a more symmetric form , 
\begin{equation}
\tilde{H}=\tilde{H}^S+\tilde{H}^T+\tilde{H}^{ST},  \label{3.9a}
\end{equation}
where 
\begin{eqnarray}
\widetilde{H}^S & = & \bar{E}_S X^{SS} +J^S\sum_{\sigma} X^{SS}n_{\sigma}, 
\nonumber \\
\widetilde{H}^T & = & \bar{E}_T \sum_\mu X^{\mu\mu}+ J^T {\bf S\cdot s} +%
\frac{J_T}{2}\sum_{\mu\sigma}X^{\mu\mu}n_{\sigma},  \nonumber \\
\widetilde{H}^{ST} & = & J^{ST}\left({\bf P}\cdot{\bf s}\right).
\label{3.14}
\end{eqnarray}
The local electron operators are defined as usual 
\begin{equation}
n_\sigma=c^\dagger_\sigma c_\sigma=\sum_{kk^{\prime}} c^\dagger_{k\sigma}
c_{k\sigma},\;\; {\bf s}=2^{-1/2}\sum_{kk^{\prime}}\sum_{\sigma\sigma^{%
\prime}} c^\dagger_{k\sigma}\hat{\tau}c_{k^{\prime}\sigma^{\prime}},
\label{3.9b}
\end{equation}
and $\hat{\tau}$ are the Pauli matrices. Moreover, the coupling constants
are, 
\begin{equation}
J^T=-\left(\frac{|W_l|^2}{\varepsilon_F-\epsilon_l}+ \frac{|W_r|^2}{%
E_r+Q-\varepsilon_F}\right),\;\;  \label{3.10a}
\end{equation}
in case (b) and 
\begin{equation}
J^T=-|W_l|^2\left(\frac{1}{\varepsilon_F-\epsilon_l}+ \frac{1}{%
E_l+Q_l-\varepsilon_F}\right),  \label{3.10b}
\end{equation}
in case (c). In both cases 
\begin{equation}
J^S=a_{ss}^2 J^T,\;\; J^{ST}=a_{ss}J^T.  \label{3.10c}
\end{equation}
This completes the derivation of the spin rotator Hamiltonian for a 
DQD hybridized with itinerant electrons.

\subsection{RG flow of coupling constants: Kondo temperature}

Due to the intermixing term $\widetilde{H}_{ST}$ in the spin Hamiltonian (%
\ref{3.14}) both triplet and singlet states are involved in the formation of
the low-energy spectrum of the DQD. Scaling equations for the coupling
constants $J_T$, $J_{ST}$ can be derived by the poor man's scaling method of
Ref. \cite{Anders70}. Neglecting the irrelevant potential scattering phase
shift and using the above mentioned procedure of integrating out the
high-energy states, a pair of scaling equations is obtained, 
\begin{equation}
\frac{dj_1}{d\ln d} = -\left[(j_1)^2+(j_2)^2\right],~~ \frac{dj_2}{d\ln d} =
-2j_1j_2 .  \label{3.13}
\end{equation}
(here $j_1=\rho_0J^T, j_2=\rho_0J^{ST}, d=\rho_0D$). If $\bar{\delta} =E_T(%
\bar{D})-E_S(\bar{D})$ is the smallest energy scale, the energy spectrum of
the DQD is quasi degenerate, and the system (\ref{3.13}) is reduced to a
single equation for the effective integral $j_+=j_1+j_2$, 
\begin{equation}
\frac{dj_+}{d\ln d} = -(j_+)^2.  \label{3.15}
\end{equation}
Then the RG flow diagram has an infinite fixed point, and the solution of
eq. (\ref{3.15}) gives the Kondo temperature 
\begin{equation}
T_{K0}=\bar{D}\exp(-1/j_+).  \label{3.15a}
\end{equation}

In the general case, the scaling behavior is more complicated. The flow
diagram still has a fixed point at infinity, but the Kondo temperature turns
out to be a sharp function of $\bar\delta.$ In the case $\bar{\delta}<0$, $|%
\bar{\delta}| \gg T_{K0}$ considered in \cite{Eto00,Pust00,KA01} the scaling
of $J^{ST}$ terminates at $D\simeq \bar{\delta}$. Then one is left with the
familiar physics of an under-screened S=1 Kondo model \cite{Noz80}. The fixed
point is still at infinite exchange coupling $J_T$, but the Kondo
temperature becomes a function of $\bar{\delta}$. It is shown in Ref. %
\onlinecite{Pust00} that a kind of universal law for $T_K(\bar{\delta})$
exists also in this limit 
\begin{equation}
T_K/T_{K0}=(T_{K0}/\bar{\delta})^\gamma,  \label{3.15b}
\end{equation}
where $\gamma$ is a numerical constant.

\subsection{Two-spin representation}

It is known \cite{Engel} that the algebra $o_{4}$ can be represented as a
direct sum of two $o_{3}$ algebras. In our case this means that one can
construct another pair of orthogonal operators 
\begin{equation}
{\bf S}_{1}=\frac{{\bf S}+{\bf P}}{2},~~{\bf S}_{2}=\frac{{\bf S}-{\bf P}}{2}%
.  \label{4.1}
\end{equation}
(see also \cite{Pust00}). In the Hubbard representation the components of
these spin vectors have a form 
\begin{eqnarray}
S_{1,2}^{+} &=&\frac{1}{\sqrt{2}}(X^{10}+X^{0\bar{1}}\pm X^{1S}\mp X^{S\bar{1%
}}),  \nonumber \\
S_{1,2}^{-} &=&\frac{1}{\sqrt{2}}(X^{01}+X^{\bar{1}0}\pm X^{S1}\mp X^{\bar{1}%
S}),  \nonumber \\
S_{z1,2} &=&\frac{1}{2}(X^{1\bar{1}}+X^{\bar{1}\bar{1}}\mp X^{0S}\mp X^{S0}).
\label{4.1a}
\end{eqnarray}
It is easy to check by direct substitution that 
\begin{equation}
S_{i}^{2}=3/4,\;X^{SS}=\frac{1}{4}-({\bf S}_{1}\cdot {\bf S}%
_{2}),\;\;\sum_{\mu }X^{\mu \mu }=\frac{3}{4}+({\bf S}_{1}\cdot {\bf S}_{2}),
\label{4.1b}
\end{equation}
The Casimir operator can be introduced as $4{\bf S}_{1}^{2}=4{\bf S}_{2}^{2}$%
.

Then substituting (\ref{4.1}) and (\ref{4.1a}) in the Hamiltonian (\ref{3.9a}%
) we rewrite it in the form 
\begin{eqnarray}
&& \tilde{H} = J ({\bf S}_1\cdot {\bf S}_2) + J_1 ({\bf S}_1\cdot {\bf s}) +
J_2 ({\bf S}_2\cdot {\bf s})+  \nonumber \\
&& J_3({\bf S}_1\cdot {\bf S}_2)\sum_\sigma
n_\sigma + const\;.  \label{4.2}
\end{eqnarray}
Here 
\begin{equation}
J=\bar{E}_T-\bar{E}_S\equiv \bar{\delta},~ J_{1,2} =J^T \pm J^{ST}, ~ J_3=%
\frac{1}{2}J_T-J_S.  \label{4.2a}
\end{equation}
Thus, as was mentioned in Ref. \onlinecite{Pust00}, the transformation (\ref
{4.1}) maps the Hamiltonian (\ref{3.9a}) on an effective two-spin Kondo
Hamiltonian plus an additional potential scattering term. However, the
physical meaning of these two spin operators differs from that in the
conventional two-site Kondo model \cite{Varma}. They only span the two $o_3$
sub-algebras of the semi-simple Lie algebra $o_4$ (see next section for
further discussion).

This kind of effective Hamiltonian appears also in other situations where
singlet and triplet states of a nanoobject are close in energy, e.g., in
vertical quantum dots \cite{Gita00,Eto00} or in conventional dots at even
occupation, provided low-lying triplet excitons are taken into account \cite
{APK,Pust00}. It was noticed in \cite{Eto00,Pust00,KA01} that the interplay
between two energy scales, i.e. the interdot singlet-triplet gap $\delta $
and the tunneling induced Kondo binding energy for triplet configuration $%
\Delta _{T}\sim \bar{D}\exp (-1/\rho _{0}J_{T})$ results in essentially
non-universal behavior of the Kondo temperature $T_{K}$ (see preceding
subsection).

\subsection{Magnetic field induced Kondo effect}

Yet another peculiar manifestation of ''hidden symmetry'' of the spin
rotator is the possible occurrence of a magnetic field induced Kondo effect.
Such possibility was discussed theoretically in Ref. \onlinecite{APK} for
the case of quantum dots formed in GaAs heterostructures and in Refs. %
\onlinecite{Pust00,Gita00,Eto00} for the case of vertical quantum dots where
the external magnetic field influences the orbital part of spatially
quantized wave functions and results in singlet-triplet level crossing. In
DQD, similar effect arises if $\bar{\delta}>0$ where the ground state of the
DQD remains a singlet in spite of the tunneling induced renormalization.
Here we re-derive the field induced Kondo effect in terms of spin rotator
representation.

\bigskip In an external magnetic field, the energy levels in $\widetilde{H}%
^{T}$ are split due to the Zeeman effect, $\widetilde{E}_{T}\rightarrow 
\widetilde{E}_{T\mu }=\widetilde{E}_{T}-\mu \delta _{Z}$. As was noticed in 
\cite{APK}, the Zeeman splitting $\delta _{Z}=g\mu _{B}B$ of the excited
triplet state compensates the energy gap $\bar{\delta}$ at a certain value
of magnetic field $B=B_{0}$. In the vicinity of this point when $\delta
-\delta _{Z}\ll \delta $ only the levels $\widetilde{E}_{T1}$ and $%
\widetilde{E}_{S}$ survive in the diagonal part ${\bar{E}}_{S}X^{SS}+{\bar{E}%
}_{T}\sum_{\mu }X^{\mu \mu }$ of the spin rotator Hamiltonian (\ref{3.14}).
Then the only renormalizable coupling parameter in the exchange Hamiltonian (%
\ref{3.14}) is $J^{ST}$. It is easily seen that the operators $%
P^{+},P^{-},(S_{z}-X^{SS})$ form an algebra $o_{3}$ in the reduced spin
space $\{S,T1\}$. In this subspace the operators $P^{+}$ and $P^{-}$ are
reduced to $\sqrt{2}X^{1S}$ and $\sqrt{2}X^{S1}$, respectively. The
operators $S^{+}\rightarrow \sqrt{2}X^{0\bar{1}}$ and $S^{-}\rightarrow 
\sqrt{2}X^{\bar{1}0}$ together with a new combination $(X^{00}-X^{\bar{1}%
\bar{1}})$ act in the subspace of excited states $\{T0,T\bar{1}\}$ divided
by the Zeeman energy from the low-energy doublet. These operators form a
complementary algebra $o_{3}$, and the direct sum of these algebras
represent a realization of the $SO(4)$ symmetry for a ''spin rotator in an
external magnetic field'' when the rotational symmetry in spin space is
broken.

As a result, the effective spin Hamiltonian (\ref{3.14}) in a subspace $\{S,
T1\}$ reduces to 
\begin{equation}
\widetilde{H}_{Z} = E_Z R_0 + J^{ST}\left({\bf R}\cdot{\bf s}\right) +H_{p}.
\label{3.16}
\end{equation}
Here $E_Z=\bar{E}_S=\bar{E}_{T1}$ is the degenerate ground state energy
level of DQD in magnetic field $B=B_0$, $H_p$ describes irrelevant potential
scattering, and the operators $R_0,$ ${\bf R}$ are 
\begin{eqnarray}
&& R_0=X^{11}+X^{SS}, R_z=X^{11}-X^{SS},  \nonumber \\
&& R^+= \sqrt{2}X^{1S},~R^- = \sqrt{2}
X^{S1}.  \label{3.17}
\end{eqnarray}
The complementary vector {\bf T} defined as 
\begin{equation}
T_z=X^{00}-X^{\bar{1}\bar{1}}, ~T^+ = \sqrt{2}X^{0\bar{1}},~T^- = \sqrt{2}X^{%
\bar{1}0},  \label{3.17a}
\end{equation}
forms a second subgroup. This vector is quenched by the magnetic field. The
spectrum of conduction electrons is also split due to Zeeman effect, but
this splitting does not affect the Kondo singularity in the tunnel current:
one simply may redefine the conduction electron energies and measure them
from the corresponding Fermi levels for spin up and down electrons \cite
{Pust00}.

Applying the poor man's scaling procedure \cite{Anders70} to the Hamiltonian
(\ref{3.16}), one comes to a scaling equation 
\begin{equation}
\frac{dj_2}{d\ln D}=-(j_2)^2  \label{3.18}
\end{equation}
with a fixed point at $j_{2}=\infty$ and the Kondo temperature $T_{KZ}=\bar{D%
}\exp(-1/j_2),$ so that 
\begin{equation}
\frac{T_K}{T_{K0}} = \exp\left(-\frac{1}{1+a_{ss}}\right).  \label{3.19}
\end{equation}

Of course, the same kind of separation is possible for a degenerate pair of
states $\widetilde{E}_S, \widetilde{E}_{T\bar{1}}$, and the corresponding
vectors ${\bf R}^{\prime}$ and ${\bf T}^{\prime}$ may be obtained from (\ref
{3.17}) and (\ref{3.17a}) by interchanging indices $1$ and $\bar{1}$ (see
Ref. \onlinecite{Pust00} for a physical realization of this situation).
\bigskip
To summarize the description of basic manifestations of spin rotator
symmetry in DQD, we considered three 
limiting cases of a spin
rotator representations depending on the physical 
situations: quasi degenerate 
state $|\bar{\delta}|\ll T_{K}$ when the
resonance properties of a DQD are determined by the full SO(4) symmetry (\ref
{3.15a}), triplet ground state $|\bar{\delta}|>T_{K}$ where the virtual
excitations to singlet state render the Kondo temperature 
to be dependent on the
initial singlet/triplet splitting (\ref{3.15b}), and magnetic-field induced
doublet ground state (\ref{3.18}) where the Kondo resonance arises in spite
of the loss of local rotational invariance. The hierarchy of Kondo
temperatures is non-universal. The maximal value of $T_{K}$ is given by $%
T_{K0}$ (\ref{3.15a}), 
from which it falls with removing singlet/triplet degeneracy 
\cite{Eto00,Pust00}. It then reaches the limiting value of $~\bar{D}\exp
(-1/j_{1})$ at large $\bar{\delta}$ where the contribution of the high lying
singlet state becomes negligibly weak, 
and one returns back to the usual $%
SU(2)$ symmetry of spin S=1 described by the $o_{3}$ algebra.

All the above results could be obtained also in the representation (\ref
{4.1a}). In this case the scaling equations should be derived for three
coupling constants $J_1,J_2,J_3$ of the spin Hamiltonian (\ref{4.2}). This
procedure is described in Ref. \onlinecite{Pust00}. 
As expected, the results are equivalent
since the scaling of $J_3$ adds nothing to the singular
behavior of the relevant parameters $J_T$ and $J_{ST}.$ The problem becomes
more complicated in case there are two sources and two drains \cite
{WW00,foot}. Then, an additional index $\alpha$ should be introduced for the
lead electrons, $c_{k\sigma}\to c_{\alpha k\sigma}$. The states with
different $\alpha$ are intermixed due to interdot tunneling and one more
operator ${\bf S}\times{\bf P}$ or ${\bf S}_1\times{\bf S}_2$ should be
introduced in the theory of spin rotator coupled to metallic leads \cite
{Pust00}.


\section{Two-center Kondo model for DQD}

It is natural to expect that in the limit of vanishing interdot coupling $V$
the tunneling through a doubly occupied DQD is defined by the individual
spins $S=1/2$ of the left and right dot, and the existence or non-existence
of resonance tunneling channel will be predetermined by the competition
between the Kondo effect for the left and right dot and indirect exchange
induced by the same tunneling. In this limit the problem is reduced to a
specific version of the two site Kondo model. The corresponding theory for 
{\it symmetric} two-site Kondo impurity in a metal was discussed, e.g., in
Refs. \onlinecite{Varma}. The question is, how this approach should be
modified in the asymmetric cases (b) and (c). On the other hand, with
increasing $V$, these two approaches should be matched, and it is
instructive to compare the description of a DQD by two fictitious spins (\ref
{4.1}) and by two real spins ${\bf S}_l$ and ${\bf S}_r$.

To study this problem we use the approach mentioned in Section 2, and
consider the DQD as a two-site center with spins 1/2 in each site within the
framework specified by the Hamiltonian (\ref{3.5a}), (\ref{3.5b}). In terms
of Hubbard operators (\ref{1.010}) for spin 1/2, this Hamiltonian is written
as $H=H_{do}+H_{t}+H_{lr}$, where 
\begin{eqnarray}
H_{do} &=&\sum_{i,\Lambda }E_{i\Lambda }X_{i}^{\Lambda \Lambda }~~~~(\Lambda
=0,\sigma ,2),  \nonumber \\
H_{t} &=&\sum_{i,k\sigma }\left[ W_{ik\sigma }\left( X_{i}^{\sigma
0}+X_{i}^{2\bar{\sigma}}\right) c_{k\sigma }+h.c.\right] ~.  \label{4.3}
\end{eqnarray}
In case (b), the states $E_{l0}$, $E_{l\sigma }=\varepsilon _{l}$, $%
E_{r\sigma }=\varepsilon _{r}$ and $E_{r2}=2\varepsilon _{r}+Q$ are involved
in the RG procedure, and the corresponding interdot tunneling Hamiltonian is
represented in the form 
\begin{equation}
H_{lr}=V\sum_{\sigma }\left( X_{l}^{\sigma 0}X_{r}^{\bar{\sigma}%
2}+h.c.\right) .\   \label{4.4}
\end{equation}
This tunneling is possible only in the singlet configuration of the DQD. We
start with eliminating the polar states $\{l0,r2\}$ that arise due to
interdot tunneling (\ref{4.4}). This procedure, known as Harris-Lange
canonical transformation \cite{Hala67}, eliminates the interdot term $H_{lr}$%
, and instead, in second order in $V$, an interdot spin-Hamiltonian emerges, 
\begin{equation}
H_{lrs}=J_{lr}\sum_{\sigma \sigma ^{\prime }}X_{l}^{\sigma \sigma ^{\prime
}}X_{r}^{\sigma ^{\prime }\sigma },  \label{4.4a}
\end{equation}
where $J_{lr}=V^{2}/\Delta _{1}$.

Like in the previous section, we should integrate out the high-energy
charged states by using the Haldane's RG procedure \cite{Hald78} for the
left dot alone, since the renormalization of the deep level $\epsilon_r$ is
negligible. However this procedure should now include the renormalization of 
$J_{lr}$ due to reduction of the conduction band. The scaling equation for $%
\varepsilon_l$ is the same as (\ref{3.6b}) for a triplet state. We rewrite
it in terms of the two-spin Hamiltonian as, 
\begin{equation}
\frac{d\varepsilon_l}{d\ln D}=\frac{\Gamma_l}{\pi},~~  \label{4.5}
\end{equation}
($\Gamma_l\equiv\Gamma_T$). The same mapping procedure as in eq. (\ref{3.1a}%
) gives the following correction to the indirect exchange coupling constant 
\begin{equation}
\widetilde{J}_{lr}=J_{lr}-\frac{\beta_1^2\Gamma_l|\delta D}{D},  \label{4.5a}
\end{equation}
and its iteration results in the second scaling equation, 
\begin{equation}
\frac{dJ_{lr}}{d\ln D}=-\beta_1^2\frac{\Gamma_l}{\pi}.~~  \label{4.5b}
\end{equation}
This procedure stops at $D=\bar{D}$ where the Schrieffer-Wolff limit for $%
\epsilon_l$ is achieved. Integrating eq. (\ref{4.5b}) from $\bar{D}$ to $D_0$
one comes to the following equation for the renormalized indirect exchange
coupling 
\[
J(\bar{D})= J(D_0)-\beta_1^2\Gamma_l\ln(D_0/\bar{D}). 
\]
Then, taking into account that $\delta=J_{lr}$ by its origin, this equation
can be rewritten in the form 
\begin{equation}
\bar{\delta}=\delta_0-\beta_1^2\Gamma_l\ln(D_0/\bar{D}).  \label{4.5c}
\end{equation}
This is the same result for the renormalized singlet-triplet excitation
energy that we have found in the preceding section. In case (c), a similar
procedure starts with eliminating the polar states generated by the interdot
tunneling term (see fig. 3c), 
\begin{equation}
H_{lr}=\sum_{\sigma} [X_l^{2\bar{\sigma}}X_r^{0\sigma} + H.c.].  \label{4.4b}
\end{equation}
In this case the Harris-Lange procedure results in the indirect exchange
Hamiltonian (\ref{4.4a}) with coupling constant $J_{lr}=V^2/\Delta_2$.
Again, only the energy level $\varepsilon_l$ is involved in the Haldane's RG
procedure. The scaling equation (\ref{4.5b}) contains on the right-hand side
the factor $\beta_2$ instead of $\beta_1$, and its solution for the
triplet/singlet level splitting gives 
\begin{equation}
\bar{\delta}=\delta_0-\beta_2^2\Gamma_l\ln(D_0/\bar{D}).  \label{4.5d}
\end{equation}
This is exactly the result obtained in Ref. {\onlinecite{KA01}. }

Next, the Schrieffer-Wolff transformation eliminates the tunnel coupling.
The operator ${\cal S}$ in (\ref{3.8}) has the form 
\begin{eqnarray}
&& {\cal S} = \sum_{k\sigma} \frac{W_l}{\varepsilon_k-\varepsilon_l}
\left(X_l^{\sigma 0}b_{k\sigma} - H.c. \right) + \nonumber \\
&& \sum_{k\sigma}\frac{W_l}{
\varepsilon_l+Q_l-\varepsilon_k} \left(X_l^{2\bar{\sigma}}b_{k\sigma} - h.c.
\right),  \label{4.6}
\end{eqnarray}
in case (b), and 
\begin{equation}
{\cal S} = \sum_{k\sigma} \frac{W_l}{\varepsilon_k-\varepsilon_l}
\left(X_l^{\sigma 0}b_{k\sigma} - h.c. \right),  \label{4.7}
\end{equation}
in case (c). As usual, in second order, the tunneling term generates an
indirect exchange between the leads and the dots. As a result, the total
spin Hamiltonian acquires the form 
\begin{equation}
H_s = \bar{J}_{lr}({\bf S}_l\cdot{\bf S}_r) + \sum_{i=l,r} J_i({\bf S}%
_i\cdot {\bf s}) +H^{\prime}.  \label{4.8}
\end{equation}
Here $\bar{J}_{lr}=\bar{\delta}$ is the renormalized singlet/triplet
excitation energy (\ref{4.5c}) or (\ref{4.5d}) in cases (b) and (c)
respectively. The components of the local spins ${\bf S}_i$ are now defined
as, 
\begin{equation}
S_i^+ =X_i^{\uparrow\downarrow},~~ S_i^- =X_i^{\downarrow\uparrow},~~ S_{iz}
=\frac{1}{2}(X_i^{\uparrow\uparrow}-X_i^{\downarrow\downarrow}).  \label{4.9}
\end{equation}
The Heisenberg-like interdot exchange (\ref{4.8}) arose in second order in $%
V $ similarly to the effective AFM exchange in a half-filled Hubbard model 
\cite{Hala67}. The indirect exchange coupling constants govern the
interaction of the conduction electrons and the local spins in the dots.
They are given by, 
\begin{equation}
J_l = \frac{|W_l|^2}{\varepsilon_F-\varepsilon_l},~~ J_r = \frac{|W_r|^2}{%
\varepsilon_l+Q -\varepsilon_F},  \label{4.10a}
\end{equation}
in case (b) and 
\begin{equation}
J_l = |W_l|^2\left( \frac{1}{\varepsilon_F-\varepsilon_l} + \frac{1}{%
\varepsilon_l+Q_l -\varepsilon_F}\right), \;\; J_r = 0,  \label{4.10b}
\end{equation}
in case (c). The last term $H^{\prime}$ in Eq. (\ref{4.8}) includes
irrelevant potential scattering terms arising in second-order
Schrieffer-Wolff transformation, and other invariants that appear in the
Hausdorff expansion (\ref{3.8}) i.e. mixed products like ${\bf S}_i \cdot [%
{\bf S}_j \times {\bf s}]$. These terms arise due to interplay between the
interdot exchange $H_{lr}$ and the tunneling $H_t$. We do not consider here
these small corrections to the main Kondo effect.

Comparing (\ref{4.8}) with (\ref{4.2}), we see that the second order terms
of the expansion reproduce the general structure of a two-spin Hamiltonian.
The interdot exchange has the same form for both representations, but there
are significant differences in the values of the coupling constants between
the leads and the local spins. In particular, the tunnel coupling between
the right dot and the leads is absent in this order in case (c), whereas in
the two-spin representation (\ref{4.1}) the coupling constants are $%
J_{1}=(2+\beta _{2}^{2})J_{l}$ and $J_{2}=\beta _{2}^{2}J_{l}$. It should be
noted, however, that nonzero coupling between the leads and the right dot
arises due to the interference between $H_{lr}$ and ${\cal S}$ in a
Schrieffer-Wolff representation, but its value differs from $J_{2}$. Thus,
from the point of view of the general $SO(4)$ symmetry of the DQD, the
two-site representation is simply one more representation of the $o_{4}$
algebra as a direct sum of two $o_{3}$ algebras. The only case when the
representations ${\bf S}_{1},{\bf S}_{2}$ and ${\bf S}_{l},{\bf S}_{r}$
coincide, is in the symmetric DQD where the admixture of excitons is ignored
and the parity is conserved for an isolated DQD. This symmetric DQD, of
course, also obeys an $SO(4)$ symmetry, but its ground state is a singlet.
The only way to activate the ''hidden symmetry'' in this case is to switch
on a strong magnetic field that compensates the exchange splitting. Then the
effective spin Hamiltonian is given in (\ref{3.16}) and a magnetic
field-induced Kondo effect arises.

\section{Concluding remarks}

It is worth making several remarks about the advantages of using alternative
approaches to analyze the physics of the quantum spin rotator. We have
presented three different ways of substituting spin 1/2 operators for the
generic operators ${\bf S}$ and ${\bf P}$ of the $SO(4)$ group. This
substitution exposes numerous connections between the approach to the Kondo
effect treating the double dot as a spin rotator and the conventional
description as a two-site Kondo problem. The traditional theory of the
two-site Kondo effect \cite{Varma} deals with a {\it symmetric} DQD, so it
is formulated in terms of even-odd spin and charge states. The effects
discussed in the present paper essentially arise only in {\it asymmetric}
situations when $J_{1}\gg J_{2}$ or $J_{l}\gg J_{r}$. Besides, we treated
conduction electrons in a single-channel approximation, whereas the even-odd
state classification of conduction electrons in a two-site Kondo model \cite
{Moust95} likes it to be a two-channel single-impurity model. Some generic
properties of the two-spin Kondo effect are, nevertheless, similar in both
limits. In particular, the competition between the on-site interactions $%
J_{i}({\bf S}_{i}\cdot {\bf s})$ and the interdot exchange $J_{ij}({\bf S}%
_{i}\cdot {\bf S}_{j})$ results in the appearance of an unstable fixed point
in the flow diagram dividing the Kondo singlet from an antiferromagnetic
singlet ground states of the system. The conventional two-site Kondo
impurity also can be classified as a ''spin rotator'', and the
singlet-triplet (even-odd) mixing is an essential part of the Kondo physics
in this case.

Nano-objects whose symmetry is more complicated than localized spins, i.e. 
{\it quantum rotors} were discussed previously in a context of the theory of
quantum phase transitions in 2D Heisenberg antiferromagnets, spin ladders
and spin glasses (see, e.g. \cite{Sach}). Quantum rotor was defined as a
spin, whose rotation is constrained to move on a surface of $M\geq 2$
-dimensional sphere. An example of array of quantum rotors is \ a double
layer system of antiferromagnetically ordered quantum spins. If the
interlayer coupling $K_{12}\left( {\bf S}_{1m}\cdot {\bf S}_{2m}\right) $ in
a site $m$ is stronger than intersite coupling $J_{mn}\left( {\bf S}%
_{1m}\cdot {\bf S}_{1n}\right) $ in a given layer, the pair of spins ${\bf S}%
_{1m},{\bf S}_{2m}$ form the quantum rotor 
${\cal S}_{m}$ with $o_{3}$
algebra and Casimir operator 
${\cal S}^{2}.$ The spin rotator is a natural
generalization of this description: in case of $M=2$ the excited triplet
state in each site $m$ can be added to a manifold, and the ladder of spin
rotors transforms into array of spin rotators.

We leave more detailed discussion of electron transport through DQD for
future communications, and discuss briefly a limiting case of a biased DQD
with $J_{r}=0$, which was realized experimentally \cite{Hoff95,Molen95}. In
case (b), this limit corresponds to an electrometer geometry . In this
configuration the right dot is isolated from the leads, but, nevertheless,
it can be used for driving the current through the left dot. In Ref. %
\onlinecite{Molen95} the driving was realized via electrostatic coupling
between the dots, and charge transfer was allowed through monitoring the
Coulomb resonance conditions. In the present case, the resonance in {\it spin%
} channel is allowed by modifying the energy of the {\it charge transfer}
exciton. To measure this effect one should choose a E-window in a plan $%
\left( V_{g}^{r},V_{g}^{l}\right) $ for symmetric DQD. At zero difference $%
V_{g}^{r}-V_{g}^{l}$ no Kondo effect should be observed. Then, increasing
this difference at given temperature $T$, one effectively changes the energy
difference $\delta _{0}$ (\ref{3.6a}) and rises the Kondo temperature $T_{K}.
$ When the regime $T\sim T_{K}$ is achieved one finds oneself in a hatched
region similar to that shown in Fig. 3a for a window $\left\{ 1,1\right\} $,
and a zero bias anomaly should appear in conductance.

In terms of the two-spin representation (\ref{4.1}), the structure of RG
scheme remains the same, and the only change is a disappearance of the
second term in equation (\ref{3.10a}) for $J_{T}$. The changes in the real
spin representation (\ref{4.9}) are more essential: $J_{r}=0$, and the Kondo
tunneling occurs only through the left dot. However, the conventional theory
of a single-site Kondo screening cannot be applied in this situation because
the interdot tunneling term $\sim \bar{J}_{lr}$ is still present in the
Hamiltonian (\ref{4.8}). If the renormalized coupling constant $\bar{J}_{lr}$
remains positive, the left spin is dynamically screened and the right spin
remains free. This model is a limiting case of under-screened spin-one
solution.

The description of the Kondo effect in terms of two fictitious spins ${\bf R}
$ and ${\bf T}$ (\ref{3.17}), (\ref{3.17a}) is another example of separation
of spin degrees of freedom into dynamically confined moment ${\bf R}$ and
unscreened moment ${\bf T}$ (see also Ref. \onlinecite{Gita00}), and this
case was realized in experiments \cite{Cobd00,Sas00} mentioned above.

We have seen that the scaling trajectory for the coupling constant $\bar{%
\delta}=\bar{J}_{lr}$ is predetermined by the bare value of the
singlet/triplet splitting $\delta_0=2\beta_1 V$, which, in turn, can be
driven by the gate voltage [see eq. (\ref{2.1})]. Thus, we see that the
Kondo tunneling channel in the left dot can be opened by softening the
charge transfer potential that is governed by the right gate voltage $V_g^r$%
, and the DQD with isolated right dot works as a "charge-spin transformer".

The limit of zero $J_r$ in case (c) was considered within the spin-rotator
approach in Ref. \onlinecite{KA01}. In terms of a full $SO(4)$ description,
the flow diagram is similar to that of a biased DQD (case b). If one would
try to describe this asymmetric DQD in terms of screening of the individual
spin ${\bf S}_l$, a problem would arise when taking into account charge
fluctuations to the state $|Ex_l\rangle$ (\ref{1.3}) at the first "Haldane"
stage of the RG procedure, because this excitation is soft by assumption, $%
\omega_{ex,l}\ll D$. In this case the source of strong correlation is, in
fact, the right dot, and the interdot tunneling is responsible for true
spin-charge separation in the DQD. The description of Kondo tunneling in
terms of the operators ${\bf S}$ and ${\bf P}$ is obviously preferable in
this case.

In conclusion, we considered here the spin excitation spectrum and the Kondo
effect in DQD from the point of view of its generic symmetry, that is, an
SO(4) symmetry of a quantum spin rotator. The properties of spin rotator
differ in many cases from those of localized spin with the same $\hat{S}^{2}$%
. In the case of triplet ground state $(\hat{S}^{2}=2)$ where cotunneling
results in under-screened Kondo effect, the presence of low-lying singlet
excitation turns the Kondo temperature to be a non-universal quantity. If the
ground state is a singlet $(\hat{S}^{2}=0)$, the Kondo effect is
nevertheless possible if one projection of the low-lying triplet excitation
is involved in electron tunneling. An alternative language for discussing
the properties of DQD is provided by a two-site Kondo model approach.
However, in spite of the overall $SO(4)$ symmetry of the problem, the
equivalence of these two approaches exists only in the case of conserved
parity (symmetric DQD). When the asymmetric charge transfer exciton is
admixed with a low-energy spin singlet, the two-site representation and the
two-spin representation of the $SO(4)$ group for a biased DQD are not
equivalent.\newline
{\bf Acknowledgment} This work is partially supported by grants from the
Israeli Science Foundations (Center of Excellence and Physics of complex
quantum dots), the US-Israel BSF grant (current instabilities in quantum
dots) and the DIP program for quantum electronics in low dimensional
systems. We benefited from discussions with I. Krive, L.W. Molenkamp and F.
M. Peeters. 
\newpage
\appendix{\bf APPENDIX} \setcounter{equation}{0} %
\renewcommand{\theequation}{A.\arabic{equation}}

The wave-functions of symmetric and asymmetric DQD occupied by one two or
three electrons are listed below, (see also \cite{Ivan}). Besides, the
tunnel matrix elements which connect the state from different charge sectors
of the DQD are presented. The eigenvalues and eigenfunctions of an isolated
neutral DQD with $N=2$ can be found by direct diagonalization of the
Hamiltonian (\ref{1.2}), (\ref{1.2a}). In a neutral configuration $\{1,1\}$
the interdot capacitive coupling is absent. Far from Coulomb resonances when
the inequalities (\ref{2.1}) are valid, expansions (\ref{1.4}) in symmetric
case (a) and (\ref{1.3}) in the asymmetric cases (b,c) give the following
equations for the coefficients $a_{ij}$ in first order of perturbation
expansion in the tunnel coupling $V$. The processes taken into account in
the mixing terms are shown by the dashed arrows in the upper panels of Fig.
1. \medskip\newline
(a) Symmetric DQD: 
\begin{equation}
a_{ee}=a_{ss}\approx 1- \beta^2,~~~~~~a_{se}=-a_{es}=\sqrt{2}\beta .
\label{A.6}
\end{equation}
(b) Biased DQD: 
\begin{eqnarray}
&& a_{ss} =  1 -\beta_1^2 - \beta_1^{\prime 2},  
a_{sl} = -a_{ls} =\sqrt{2}
\beta_1^{\prime},  \nonumber  \\
&& a_{sr} = -a_{rs} =\sqrt{2}\beta_1,  \nonumber \\
&& a_{ll} =  1-\beta_1^{\prime 2},  a_{rr} = 1-\beta_1^2 .  
\label{A.8}
\end{eqnarray}
Here $\beta_1^{\prime}=V/(Q+\Delta)$. We assume that $\beta_1^{\prime}\ll
\beta_1$ and neglect the terms $\sim \beta_1^{\prime}$ in our calculations.
\medskip\newline
(c) Asymmetric DQD: 
\begin{eqnarray}
&& a_{ss} =  1 -\beta^{2}_2 - \beta^{\prime 2}_2, a_{sl} = -a_{ls} =\sqrt{2
}\beta_2,  \nonumber \\
&& a_{sr} = -a_{rs} =\sqrt{2}\beta^\prime_2,  \nonumber \\
&& a_{ll} = 1-\beta_2^{2},  a_{rr} = 1-\beta^{\prime2}_2 .  \label{A.10}
\end{eqnarray}
Here $\beta_2^\prime=V/(Q_r-\Delta)\ll \beta_2$, and we neglect the
corresponding contributions as well.

To complete the enumeration of states involved in the tunneling Hamiltonian (%
\ref{1.1}) one should define the charge states of DQD which arise in a
process of electron tunneling between the DQD and metallic leads. We are
interested in one-electron tunneling, so the states with one and three
electrons in DQD, $N=1,3$, should be taken into account. \medskip\newline
(a) Symmetric DQD: \medskip\newline
One-electron states are even and odd combinations of electronic wave
functions belonging to the left and right well. The same is valid for the
three-electron states which in fact are the hole analogs of one-electron
states. 
\begin{eqnarray}
|1e,\sigma\rangle & = & \frac{1}{\sqrt{2}}\left(d^\dagger_{l\sigma}+
d^\dagger_{r\sigma}\right)|0 \rangle ,  \nonumber \\
|1o,\sigma\rangle & = & \frac{1}{\sqrt{2}}\left(d^\dagger_{l\sigma}-
d^\dagger_{r\sigma}\right)|0 \rangle , \label{A.12} \\
|3e,\sigma\rangle & = & \frac{1}{\sqrt{2}}\left(
d^\dagger_{l\sigma}d^\dagger_{r\downarrow}d^\dagger_{r\uparrow}+
d^\dagger_{r\sigma}d^\dagger_{l\downarrow}d^\dagger_{l\uparrow}\right)|0%
\rangle , \nonumber \\
|3o,\sigma\rangle & = & \frac{1}{\sqrt{2}}\left(
d^\dagger_{l\sigma}d^\dagger_{r\downarrow}d^\dagger_{r\uparrow}-
d^\dagger_{r\sigma}d^\dagger_{l\downarrow}d^\dagger_{l\uparrow}\right)|0%
\rangle . \nonumber
\end{eqnarray}
(b,c) Asymmetric DQD: \medskip\newline
In this case the DQD is "polarized" both in negatively and positively
charged states. The one-electron wave functions are the same in cases (b)
and (c) 
\begin{eqnarray}
|1a,\sigma\rangle & = & \left(\sqrt{1-\alpha^2}\;d^\dagger_{l\sigma}-
\alpha\; d^\dagger_{r\sigma}\right)|0\rangle , \nonumber \\
|1b,\sigma\rangle & = & \left(\alpha\; d^\dagger_{l\sigma}+ \sqrt{1-\alpha^2}%
\;d^\dagger_{r\sigma}\right)|0\rangle  \label{A.13}
\end{eqnarray}
($\alpha=V/\Delta$). The corresponding energy levels are 
\begin{equation}
E_{1a}=\varepsilon_l+\alpha V,~~ E_{1b}=\varepsilon_r -\alpha V,~~
\label{A.13a}
\end{equation}
The three-electron wave functions are represented by expressions 
\begin{eqnarray}
|3b,\sigma\rangle & = & \left( \sqrt{1-\alpha^2}\;
d^\dagger_{l\sigma}d^\dagger_{r\downarrow}d^\dagger_{r\uparrow} + \alpha\;
d^\dagger_{r\sigma}d^\dagger_{l\downarrow}d^\dagger_{l\uparrow}\right)|0%
\rangle , \nonumber \\
|3a,\sigma\rangle & = & \left(-\alpha\;
d^\dagger_{l\sigma}d^\dagger_{r\downarrow}d^\dagger_{r\uparrow} + \sqrt{%
1-\alpha^2}\;
d^\dagger_{r\sigma}d^\dagger_{l\downarrow}d^\dagger_{l\uparrow}
\right)|0\rangle  \label{A.14}
\end{eqnarray}
in case (b), and 
\begin{eqnarray}
|3b,\sigma\rangle & = & \left( \sqrt{1-\alpha^{\prime2}}\;
d^\dagger_{r\sigma}d^\dagger_{l\downarrow}d^\dagger_{l\uparrow} +
\alpha^\prime\;
d^\dagger_{l\sigma}d^\dagger_{r\downarrow}d^\dagger_{r\uparrow}\right)|0%
\rangle,  \nonumber \\
|3a,\sigma\rangle & = & \left(-\alpha^\prime\;
d^\dagger_{r\sigma}d^\dagger_{l\downarrow}d^\dagger_{l\uparrow} + \sqrt{%
1-\alpha^{\prime2}}\;
d^\dagger_{l\sigma}d^\dagger_{r\downarrow}d^\dagger_{r\uparrow}
\right)|0\rangle  \label{A.15}
\end{eqnarray}
in case (c). Here $\alpha^\prime =V/(Q_r-Q_l-\Delta).$ The eigen-levels are
given by the following equations 
\begin{equation}
E_{3b}=2\varepsilon_r+Q-\alpha V,~~E_{3a}=2\varepsilon_l+Q+\alpha V
\label{A.16}
\end{equation}
in case (b) and 
\begin{equation}
E_{3b}=2\varepsilon_l+Q_l-\alpha^\prime V,~~ E_{3a}=2\varepsilon_r+Q_r
+\alpha^\prime V  \label{A.17}
\end{equation}
in case (c).

The tunneling matrix elements in the Hamiltonian $H_t$ (\ref{1.03}) include
states from different charge sectors $N(\nu_l,\nu_r)$ of the dot Hamiltonian 
$H_d$ (\ref{1.01}). In the presence of an interdot coupling $V$, and at
nonzero bias potential $V_g^l- V_g^r >0,$ the numbers $\nu_l,\nu_r$ are
non-integer, and the tunneling transparencies of the left and right dot are
different even if $W_l=W_r$ (case b). In case (c), the tunneling barrier
between the leads and the right dot is wider, and one can assume that $%
W_r<W_l$, so that the symmetry is even stronger. Consideration of the
asymmetric configurations in case (b), we note that the expansion
coefficients in eq. (\ref{1.3}) for the two electron states $|\Lambda\rangle$
are such that $a_{ss}\gg a_{sl}, a_{sr}$ (see eq. (\ref{A.8})). \ The tunnel
matrix elements which define the dominant contributions to the RG equations (%
\ref{3.1b}) are 
\begin{eqnarray}
&& W_{qu\sigma}^{TO,1\bar{\sigma}} =  \frac{1}{\sqrt{2}}w_l, 
\nonumber \\
&& W_{qu\pm}^{T\pm,1\pm}=w_l,~ W_{qd\sigma}^{TO,3\bar{\sigma}}=\frac{\sigma}{%
\sqrt{2}}w_r, \nonumber \\ 
&& W_{qd\pm}^{T\pm,3\pm}=\sigma w_r,  \nonumber \\
&& W_{qu\sigma}^{S,1\bar{\sigma}}  = \frac{1}{\sqrt{2}}\sigma 
a_{ss}w_l, \nonumber \\
&& W_{qd\sigma}^{S,3\bar{\sigma}}=\frac{1}{\sqrt{2}}a_{ss}w_r  \label{3.4}
\end{eqnarray}
(here $w_l=\sqrt{1-\alpha^2}W_l$). Similar equations can be derived in case
(c), where the wave-functions of the virtual charged states $%
|1b\sigma\rangle $ and $|3b\sigma\rangle$ are given by eqs (\ref{A.13}), (%
\ref{A.15}). Now instead of (\ref{3.4}) one has 
\begin{eqnarray}
&& W_{qu\sigma}^{TO,1\bar{\sigma}} =  \frac{1}{\sqrt{2}}w_l,
W_{qu\pm}^{T\pm,1\pm}=w_l, \nonumber \\ 
&& W_{qd\sigma}^{TO,3\bar{\sigma}}=\frac{\sigma}{%
\sqrt{2}}w^\prime_l,~ W_{qd\pm}^{T\pm,3\pm}=\sigma w_r,  \nonumber \\
&& W_{qu\sigma}^{S,1\bar{\sigma}}  =  \frac{1}{\sqrt{2}}\sigma a_{ss}w_l,~
W_{qd\sigma}^{S,3\bar{\sigma}}=\frac{1}{\sqrt{2}}a_{ss}w^\prime_l
\label{3.7}
\end{eqnarray}
where $w^\prime_l=\sqrt{1-\alpha^{\prime2}}w_l$.


\end{multicols}

\begin{references}
\bibitem{foot}  Although in experimental setups with parallel geometry each
of the dots is coupled with its own pair of leads we consider here the case
of single source and single drain reservoir. The generalization of the
theory for two channel geometry is straightforward (see Ref. %
\onlinecite{Pust00})
\bibitem{transf}  In general case of non-integer $\nu_{l,r}$ the interdot
Coulomb and exchange interaction arises because of the charge transfer in
asymmetric configurations (b) and (c). This interaction is $\sim \beta_j^4$ (%
$j=1,2$, respectively), and we neglect it in further calculations.
\end{references}
\end{document}